\documentclass{article}
\usepackage{geometry}
\date{}

\usepackage{authblk}
\usepackage{lineno}
\usepackage{float}
\usepackage{amsmath,amssymb,graphicx,color}
\usepackage{bm}
\usepackage{txfonts}
\usepackage[T1]{fontenc}
\usepackage{soul}
\usepackage{braket,commath}
\usepackage{upgreek}
\usepackage{caption}
\usepackage[numbers,sort&compress]{natbib}

\soulregister\ref{7}
\soulregister\eqref{7}
\soulregister\cite{7}
\soulregister\onlinecite{7}

\begin{document}

\title{\textbf{\LARGE Vectorial microlasers with designable topological charges based on M\"{o}bius-like correspondence in quasi-BICs}}

\author[1,6]{Xinhao Wang}
\author[1,6]{Zhaochen Wu}
\author[1,5,6,*]{Jiajun Wang}
\author[1,2,3,4,5,*]{Lei Shi}
\author[1,2,3,4,5,*]{Jian Zi}

\affil[1]{\raggedright State Key Laboratory of Surface Physics, Key Laboratory of Micro- and Nano-Photonic Structures (Ministry of Education) and Department of Physics, Fudan University, Shanghai 200433, China.}
\affil[2]{Institute for Nanoelectronic devices and Quantum computing, Fudan University, Shanghai 200438, China.}
\affil[3]{Collaborative Innovation Center of Advanced Microstructures, Nanjing University, Nanjing 210093, China.}
\affil[4]{Shanghai Research Center for Quantum Sciences, Shanghai 201315, China.}
\affil[5]{Shanghai Key Laboratory of Metasurfaces for Light Manipulation, Fudan University, Shanghai 200433, China.}
\affil[6]{These authors contributed equally: Xinhao Wang, Zhaochen Wu, Jiajun Wang.}
\affil[*]{e-mail: jiajunwang@fudan.edu.cn; lshi@fudan.edu.cn; jzi@fudan.edu.cn}

\maketitle

\large
\section*{Abstract}

The ability to control topological properties of laser emission represents a fundamental advancement in photonic technology. Achieving topological laser in a single compact photonic structure is crucial for device integration and miniaturization but faces significant challenges for designing both the high-quality (high-Q) mode and radiative topological configurations. Recently, bound states in the continuum (BICs), as extraordinary states possessing both ultrahigh Q factors and polarization topological charges, have been demonstrated as a promising platform for compact topological lasers. However, as the cornerstone of BIC lasing's non-trivial properties, topological charges of BICs are protected by real-space structural symmetries, which simultaneously impose fundamental limitations that hinder their designability of lasing topological charges. Here, we propose and experimentally demonstrate a compound cavity design method based on the M\"{o}bius-like correspondence in quasi-BICs (q-BICs), by which compact vectorial microlasers with designable topological charges can be realized.  We reveal the hidden connection between real-space symmetry breaking and eigen-polarizations of q-BICs from the triangular photonic crystal (PhC) slab, manifesting as a M\"{o}bius-like correspondence. By splicing PhC slab sectors utilizing this M\"{o}bius-like correspondence, we establish a one-to-one correspondence between compound cavities and their lasing topological charges. Vectorial lasing with designable topological charges from $-5$ to $+5$ were experimentally realized. Our work establishes a novel BIC-based platform that enables designable topological lasing, providing a promising route toward compact topological sources.

\section*{Introduction}

The presence of non-trivial topological structures in lasing profiles endows optical fields with rich properties and functionalities~\cite{zhan2009cylindrical,chen2018vectorial,shen2019optical,forbes2019structured,ni2021multidimensional,forbes2021structured,forbes2024orbital}, enabling many applications from metrology~\cite{fang2021vectorial}, microscopy~\cite{dorn2003sharper,liu2022super}, optical manipulation to quantum information processing~\cite{parigi2015storage,ndagano2017characterizing,sit2017high} and light-matter interactions~\cite{sederberg2020vectorized,el2021sensitive}, etc. Directly implementing lasing with non-trivial topological structures through single compact photonic structures represents a highly desirable yet challenging goal in modern optic, as it would substantially advance the integration and miniaturization of photonic devices while maintaining their complex functionalities. Progress has been made through various approaches like Dirac-vortex cavities~\cite{gao2020dirac,yang2022topological,han2023photonic,ma2023room}, compound cavities with metasurfaces~\cite{sroor2020high,piccardo2022vortex,ni2022spin} and other topological mode cavities~\cite{hwang2024vortex,yang2020spin}, etc~\cite{miao2016orbital,zhang2020tunable,chen2025observation}; yet, substantial challenges persist in achieving non-trivial lasing profiles with designability of the topological charges and single-compact-structure implementation. For instance, while Dirac-vortex cavities have demonstrated the possibility of vectorial lasing with polarization topological charges through topological mid-gap modes, the correspondence between cavity structures and topological charges remains ambiguous~\cite{gao2020dirac}. Meanwhile, conventional resonant metasurfaces, limited by their lack of high-quality (High-Q) modes, typically require additional cavities to achieve topological charge lasing~\cite{sroor2020high,piccardo2022vortex,ni2022spin}. For most existing topological microlasers based on cavity modes, the achievable topological charges are typically determined through numerical calculation after cavity design, making the purposeful design of desired topological charges particularly challenging, with realized topological charges confined to low order. These limitations not only point to further directions for those research areas but also highlight the crucial requirements for exploring novel design methods of topological microlasers.

In the exploration of novel topological microlaser designs, bound states in the continuum (BICs) have emerged as a promising platform due to their unique properties revealed in recent studies of topological photonics~\cite{hsu2016bound,kang2023applications,huang2023resonant,wang2024optical}. BICs have been found to exhibit unexpected polarization vortices in momentum space, offering vectorial topological configurations in photonic bands~\cite{zhen2014topological,zhang2018observation,doeleman2018experimental}. These counterintuitive momentum-space vortices are supported by real-space periodic structures like photonic crystal (PhC) slabs, manifesting as winding polarization states of eigen optical modes surrounding BICs. By accumulating the winding angle in polarization vortices, quantized numbers can be defined as topological charges of BICs~\cite{hsu2016bound}. The non-trivial topological configurations of BICs also enable another prominent property, i.e., infinite Q factors. Building on these properties, BICs have been applied in resonant and radiative systems to generate light fields with non-trivial characteristics, such as vortex beams based on momentum-space Pancharatnam-Berry phase~\cite{wang2020generating}, Bose-Einstein condensation~\cite{ardizzone2022polariton,gianfrate2024reconfigurable} and vectorial lasing with topological charges~\cite{kodigala2017lasing,huang2020ultrafast,hwang2021ultralow,sang2022topological,chen2023compact,wang2025inherent,cui2025ultracompact}.

Microlaser design based on the BIC concept represents a highly valuable method for compact topological lasing. The non-trivial lasing profiles are dominated by momentum-space topological vortex configurations of BICs, which are protected by real-space structural symmetries~\cite{hsu2016bound,wang2024optical}. However, these symmetries fundamentally constrain the allowable topological charges of BICs, as only specific vortex configurations are permitted under given symmetry operations. This restriction on topological charges consequently limits the controllability and designability of current BIC-based microlaser systems. Breaking original structures' symmetry is necessary for the goal of purposefully designing different topological charges. While, the fundamental inquiry of how to construct topological laser modes from symmetry-breaking BIC systems remains largely unexplored.

In this work, we introduce a compound cavity design strategy for realizing vectorial microlasers with designable topological charges by leveraging M\"{o}bius-like correspondence in quasi-BICs (q-BICs). Here, q-BICs are high-Q optical modes that arise from symmetry-protected BICs through slight symmetry breaking, with their eigen-polarizations intrinsically linked to the symmetry-breaking parameters. We explored the hidden connection between structural symmetry breaking and eigen-polarizations of q-BICs from a high-order BIC with $-2$ topological charge in the triangular PhC slab, which manifests as a M\"{o}bius-like correspondence. Guided by this correspondence, multiple q-BIC PhC slabs with distinct structural parameters and eigen-polarizations are spliced in an ordered angular sequence to form compound cavities supporting vectorial topological lasing modes. This cavity construction concept based on M\"{o}bius-like correspondence establishes a one-to-one correspondence between cavity configuration and lasing topological charge. Experimentally, vectorial lasing with designable topological charges ranging from $-5$ to $+5$ were realized.

\section*{Results}
\subsection*{Concept}

\begin{figure}[H]
 \centering
  \includegraphics[scale=1]{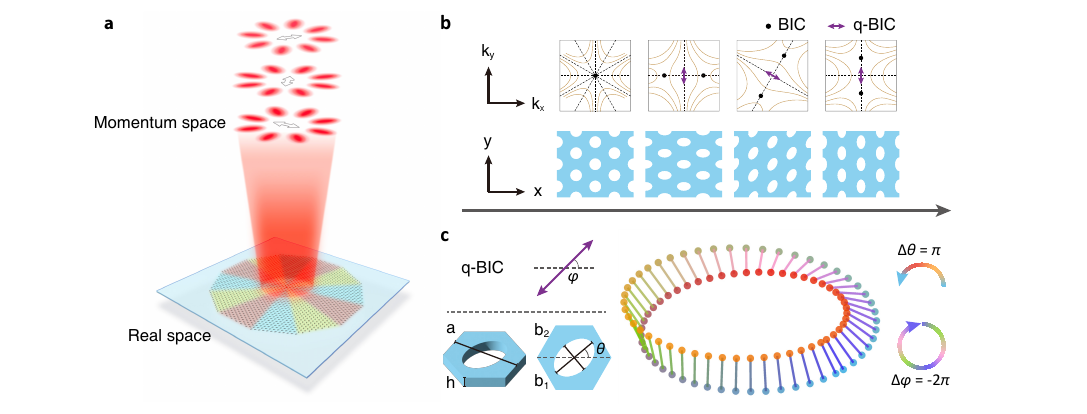}
  \caption{$\textbar$ \textbf{Principle of the vectorial microlaser with designable topological charge.} \textbf{a}, Schematic of a vectorial microlaser with $-4$ topological charge. The microlaser cavity is constructed by arranging three types of PhC slabs which support q-BICs of different eigen-polarizations. \textbf{b}, Schematic, relationship between momentum-space polarization configurations and PhC structures. Lower panel shows top view of PhC slabs. Upper panel shows the corresponding polarization distribution, in which dashed lines mark the mirror planes, and the solid lines represent the continuous evolution of the major axes of the eigen-polarization states across different optical modes. \textbf{c}, Left panel, definition of the polarization orientation of q-BIC ($\varphi$, ranging from $-\pi/2$ to $\pi/2$) and the rotation angle of elliptical hole ($\theta$, ranging from 0 to $\pi$). Right panel, the correspondence between the PhC and q-BIC manifests as a M\"{o}bius-strip-like representation. The spheres embedded at the edge of the M\"{o}bius strip refers to PhC of different rotation angle $\theta$, and the surface of the M\"{o}bius strip (the lines between the two spheres) refers to the polarization orientation $\varphi$ of the q-BIC. As $\theta$ evolves from 0 to $\pi$ clockwise or counterclockwise ($\left|\Delta\theta\right|=\pi$), corresponding q-BIC completes a closed loop (two polarization cycles) along the edge of the M\"{o}bius strip ($\left|\Delta\varphi\right|=2\pi$).} 
  \label{Fig:1}
\end{figure}

To introduce the basic concept, we begin with the continuous evolution of q-BICs under symmetry-breaking parameter manipulating. The first column of Fig. \ref{Fig:1}\textbf{b} shows the top view of a triangular-lattice PhC slab and the corresponding polarization distribution in momentum space in a photonic band. A symmetry-protected BIC is located at the momentum-space center ($\Gamma$ point). With C$_{6}$ rotational symmetry, the BIC is enabled with a topological charge of $-2$. When slightly breaking the C$_{6}$ symmetry to C$_{2}$ symmetry via changing the circular holes to be elliptic holes, momentum-space polarization configurations vary and at $\Gamma$ point the BIC evolves to a q-BIC~\cite{koshelev2018asymmetric,liu2019high}. Under slight symmetry breaking, the at-$\Gamma$ q-BIC still has high Q factor and carries linear polarization~\cite{liu2019high,yoda2020generation,wang2022realizing}. The orientation of linear polarization in momentum space ($\varphi$ in Fig. \ref{Fig:1}\textbf{c}) shows close connection with the symmetry breaking, more precisely, the orientation of elliptic holes in real space ($\theta$ in Fig. \ref{Fig:1}\textbf{c}). The q-BIC’s polarization orientation $\varphi$ rotates continuously along with the symmetry breaking orientation $\theta$, as shown in Fig. \ref{Fig:1}\textbf{b}. Notably, in the second and fourth columns of Fig. \ref{Fig:1}\textbf{b}, we can see two types of PhC slabs with different symmetry breaking orientation $\theta$ can have q-BICs with the same polarization orientation $\varphi$. The continuous parameter relationship between the real-space symmetry-breaking structure and the eigen-polarization of its supported q-BIC can be represented by the M\"{o}bius-like correspondence shown in Fig. \ref{Fig:1}\textbf{c}. As the angle $\theta$ evolves from $0$ to $\pi$, the q-BIC’s polarization orientation $\varphi$ undergoes a complete cycle of $2\pi$, which can be described as a closed loop on the M\"{o}bius strip. 

In short, the M\"{o}bius-like correspondence between the structural symmetry breaking and eigen-polarization of q-BIC establishes a foundational framework for achieving designable topological charges with compound microlaser cavity. Based on the elucidated design and functionality precursor in Figs. \ref{Fig:1}\textbf{b} and \textbf{c}, the general methodology for realizing vectorial lasing with designable topological charge was schematically illustrated in Fig. \ref{Fig:1}\textbf{a}. The compound microlaser cavity is constructed by splicing different regions of PhC designs on a single slab, with a definite structural center. These colored PhC slabs, all sharing the same triangular lattice, are modified in terms of the orientations of the elliptical holes to support q-BICs with distinct linear polarizations. The arrangement principle is based on the polarization continuity enabled by the M\"{o}bius-like correspondence, to finally form a compound cavity supporting topological charge lasing mode. By intentionally designing the interior arrangement for the compound cavity, we can purposefully control the absolute value and sign of the lasing topological charge.

\subsection*{Experimental realization}

To directly demonstrate the M\"{o}bius-like correspondence in q-BICs, we performed lasing experiments via PhC slabs with different symmetry-breaking parameters. The designed PhC slabs are fabricated by etching periodic elliptical holes in a silicon nitride (Si$_3$N$_4$, refractive index $\sim$2) film on the optical silica substrate, with the unit cell shown in Fig. \ref{Fig:1}\textbf{c}. The lattice constant $a$ is 634 nm, and Si$_3$N$_4$ thickness $t$ is 140 nm. For the etched elliptical holes, the major/minor axis lengths are fixed at $b_1 = 480$ nm and $b_2 = 320$ nm.

\begin{figure}[H]
 \centering
  \includegraphics[scale=1]{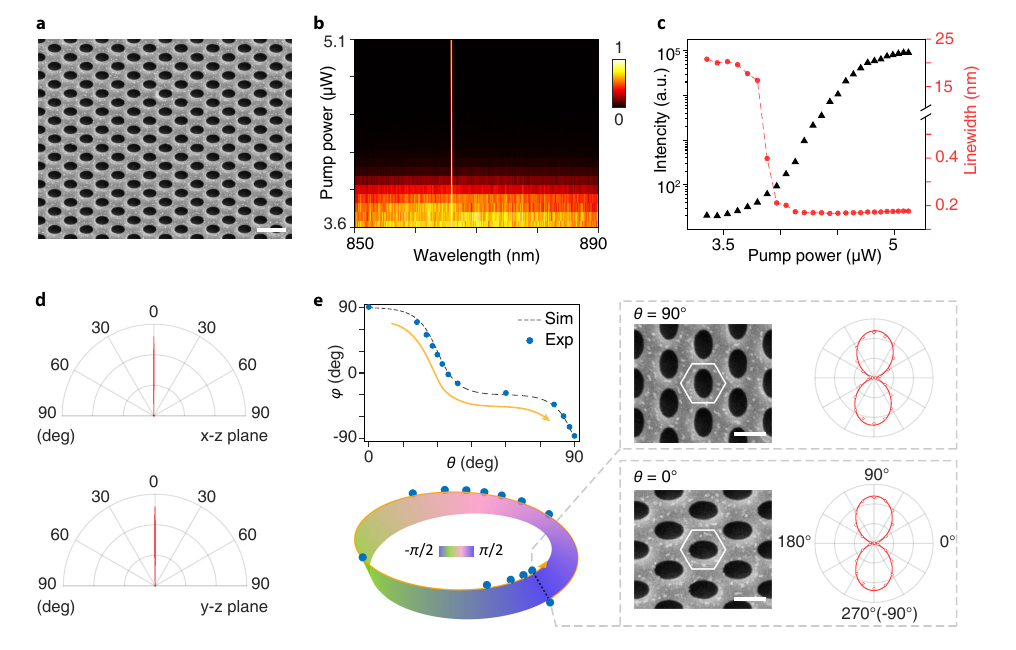}
  \caption{\textbf{$\textbar$ Lasing properties via q-BIC and measured M\"{o}bius-like correspondence.} \textbf{a}, SEM images of the fabricated PhC slab with $\theta = 0^{\circ}$. Scale bar, 1 $\upmu$m. \textbf{b}, Normalized PL spectra at varying pump powers. \textbf{c}, Light–light curve (in black) and linewidth evolution curve (in red) as functions of pump power for the lasing mode. \textbf{d}, Measured directivity of the emission above the lasing threshold in x-z and y-z planes. \textbf{e}, Left, lasing emission polarizations $\varphi$ measured as function of the structural parameter $\theta$. Experimental results (blue circles) show good agreement with the simulated polarization of the quasi-BIC (grey dashed line) in the upper panel, confirming the M\"{o}bius-like correspondence (bottom). The color of the M\"{o}bius strip describes the polarization orientation $\varphi$. Right, SEM images and experimental measured lasing emission polarizations of the PhC slab with $\theta = 0^{\circ}$ (top) and $\theta = 90^{\circ}$ (bottom), corresponding to the two end points of a tangent line on the M\"{o}bius strip (black dashed line). Scale bar, 500 nm.}
  \label{Fig:2}
\end{figure}

To give a specific example in detail, we simulated and experimentally characterized one PhC slab ($\theta = 0^{\circ}$). In simulations, the PhC slabs are embodied in the optical silica environment (refractive index $\sim$1.45). For practical experiments, PhC slabs are immersed in dimethyl sulfoxide (DMSO) to match the refractive index of optical silica. As shown in Supplementary Information Section 3, we firstly analyzed the designed photonic band (blue color) and the momentum-space properties. In this case, the at-$\Gamma$ q-BIC carries a y-direction eigen polarization ($\varphi = 90^{\circ}$) and has a high Q factor. This designed PhC slab was fabricated by electron beam lithography process, and Fig. \ref{Fig:2}\textbf{a} shows corresponding scanning electron microscopy (SEM) image. The photonic band and q-BIC are directly observed and characterized by the polarization-analyzed momentum-space spectroscopy measurement system, as shown in Supplementary Information Section 3. To achieve q-BIC lasing, IR-140 dye molecules was chosen as the gain medium and was dissolved into the DMSO overlaid on the PhC slabs. A femtosecond pulse laser was applied as the pumping source to validate the q-BIC lasing behaviors. Detailed fabrication and optical measurement procedures are provided in Methods and Supplementary Information.

Fig. \ref{Fig:2}\textbf{b} displays the normalized photoluminescence (PL) spectra at varying pump intensities, clearly revealing a phase transition from spontaneous emission to lasing emission. The corresponding light-light curve and linewidth evolution curve are extracted from the PL spectra, as shown in Fig. \ref{Fig:2}\textbf{c}. The observed S-shaped curve and dramatic reduction in linewidth are characteristic of lasing threshold behavior, further confirming the transition to lasing emission. Above the threshold, the lasing emission matches the q-BIC mode both in wavelength and radiation direction. As shown in Fig. \ref{Fig:2}\textbf{d}, the lasing emission exhibits high directivity along the z-axis (normal direction, corresponding to the $\Gamma$ point in momentum space) with radiation angles of $\pm0.6^{\circ}$ in the x-z plane and $\pm0.8^{\circ}$ in the y-z plane, validating the q-BIC lasing. The lower right panel of Fig. \ref{Fig:2}\textbf{e} exhibits the measured polarization of the q-BIC lasing ($\varphi = 90^{\circ}$), agreeing well with the linear polarization in the simulation.

A series of the PhC slabs with different rotation angles $\theta$ were further fabricated to directly map the M\"{o}bius-like correspondence between eigen-polarizations of q-BICs and the structural symmetry breaking via the lasing measurements. We measured their emission polarization orientations above threshold and plotted them as a function of $\theta$ in the upper left panel of Fig. \ref{Fig:2}\textbf{e}. Detailed SEM images and measured lasing emission polarizations are shown in the right panels of Fig. \ref{Fig:2}\textbf{e} and Supplementary Information Section 5. The experimental data align well with the simulations and exhibit a M\"{o}bius-like correspondence (left panel of Fig. \ref{Fig:2}\textbf{e}) following the principles introduced in Fig. \ref{Fig:1}\textbf{c}. As indicated by the orange arrows, the experimental results trace half the path along the edge of the M\"{o}bius strip, achieving full linear polarization coverage on equator of the Poincar\'e sphere. Notably, for PhC slabs of $\theta = 0^{\circ}$ and $90^{\circ}$, their polarization orientations coincide and are located at the two endpoints of a tangent line (black dashed line), embodying the M\"{o}bius-like correspondence in q-BICs. Exploiting the mirror symmetry of the structures, the remaining half path can be extrapolated from the current results, together forming a complete M\"{o}bius-like correspondence in Fig. \ref{Fig:1}\textbf{c}.

\begin{figure}[H]
 \centering
  \includegraphics[scale=1]{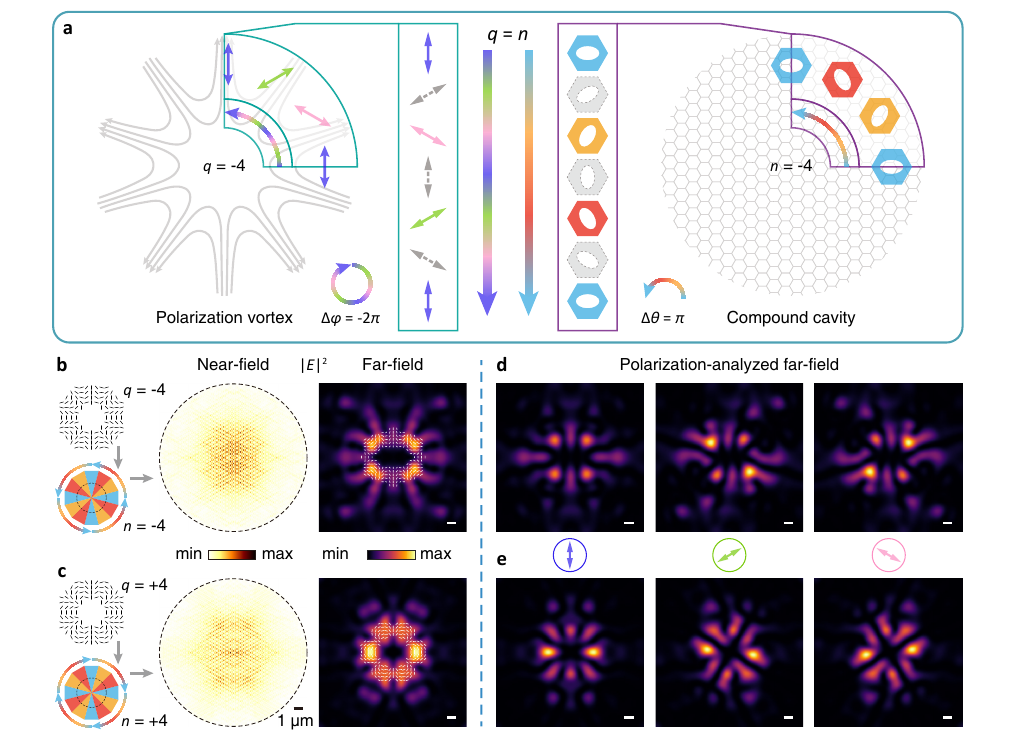}
  \caption{$\textbar$ \textbf{Construction of compound cavities based on M\"{o}bius-like correspondence.} \textbf{a}, Schematic of the construction principle. Left, illustration of the polarization vortex with $-4$ topological charge. Right, schematic of the constructed compound cavity consisting of angularly repeating sections. The repeating section is composed of three types of q-BIC PhC slabs with $\theta = 0^{\circ}$ (blue), $60^{\circ}$ (yellow), $120^{\circ}$ (orange). The middle panel shows the M\"{o}bius-like correspondence, where grey dashed polarizations and unit cells of PhC are plotted to show the intermediate parts along the minimum parameter path. \textbf{b–c}, Construction of compound cavities for vectorial lasing with topological charges of $-4$ (\textbf{b}) and $+4$ (\textbf{c}). Left: schematic of the targeted polarization vortices and corresponding cavity design. Right: simulated near-field and far-field profiles of the cavity modes. The far-field polarization major axis distributions agree well with the targeted polarization vortices. Scale bar, $1^{\circ}$. \textbf{d–e}, Simulated polarization-analyzed far-field profiles of the cavity modes $q = -4$ (\textbf{d}) and $q = +4$ (\textbf{e}) . Colored arrows denote the analyzed polarization directions. Scale bar, $1^{\circ}$.}
  \label{Fig:3}
\end{figure}

With the M\"{o}bius-like correspondence, we demonstrated the construction method of compound microlaser cavities supporting vectorial lasing mode with designable topological charges. To clarify the general design principle of lasing topological charge, we begin with the example of a vectorial lasing mode with a topological charge of $-4$, as illustrated in Fig. \ref{Fig:3}\textbf{a}. The topological charge is characterized from far-field polarization vector distribution, which exhibits a vortex configuration with a singularity at the center~\cite{hwang2024vortex}.

Specifically, the topological charge $q$ is determined by the total winding angle of the polarization orientation $\varphi$ along a closed loop traced in the counterclockwise direction~\cite{hsu2016bound,wang2024optical}:
\begin{equation}\label{eq:1}
q = \frac{1}{2\pi}\oint_{C} \frac{\partial \varphi}{\partial \psi} \mathrm{d}\psi.
\end{equation}
Here, $\psi$ denotes the azimuthal angle within the two-dimensional plane of the vortex. From previous results, we can see the elliptic hole's rotation of $\pi$ refers to a polarization winding of $2\pi$ in the discovered M\"{o}bius-like correspondence, reflecting an absolute topological charge of $1$, while the sign of the topological charge can be controlled by adjusting the direction of the polarization rotation. Then, for the considered vectorial lasing mode with $-4$ topological charge (left panel in Fig. \ref{Fig:3}\textbf{a}), the polarization distribution can be decomposed into four angularly arranged repeating sections, each contributing a topological charge of $-1$ and highlighted by a cyan sector. Guided by the M\"{o}bius-like correspondence shown in the middle panel of Fig. \ref{Fig:3}\textbf{a}, we translate the targeted polarization states (left panel) into the structural parameters of q-BIC PhC sections in real-space counterpart (right panel). To support the targeted topological polarization vortex distribution, three types of q-BIC PhC slabs each with a different rotation angle, are arranged along the angular direction to form the real-space repeating section, highlighted by the purple sector (right panel in Fig. \ref{Fig:3}\textbf{a}). With the consideration for the minimum path of the M\"{o}bius-like correspondence, just three types of PhC slabs can ensure the required polarization variations in the designed structure. In the middle panel, grey dashed arrows and hexagons indicate the intermediate q-BIC polarizations and PhC unit cells along the minimal path on the M\"{o}bius-like correspondence.

Once the repeating section is established, constructing topological charges becomes straightforward. By angularly repeating the real-space section $n$ times, a compound cavity is formed that supports a vectorial lasing mode carrying the targeted polarization vortex with topological charge $q$. The sign of topological charge can be controlled by the angular arrangement of PhC slabs: clockwise assembly yields positive $q$, while counterclockwise assembly yields negative $q$. For clarity, we define the sign of the repetition number $n$ to indicate the arrangement direction, with $n>0$ corresponding to clockwise and $n<0$ to counterclockwise. Consequently, our construction method exhibits a unique feature: a one-to-one correspondence between the compound cavities and their lasing topological charges, explicitly given by
\begin{equation}\label{eq:2}
n = q.
\end{equation}

To validate the construction strategy for compound cavity, we further consider the cases of topological charges $-4$ and $+4$. Starting from the targeted topological charges of the nontrivial polarization distributions, as shown in the left panels of Figs. \ref{Fig:3}\textbf{b} and \textbf{c}, two compound cavities were accordingly designed by angularly repeating the real-space sections $-4$ and $+4$ times, respectively. The angular arrangement of the q-BIC PhC slabs constructs a compound cavity and supports a cavity mode localized at the structure centers for vectorial lasing, as confirmed by the simulated near-field profiles in the right panels of Figs. \ref{Fig:3}\textbf{b} and \textbf{c}. These modes carry polarization vortices in the far field, where the extracted polarization major axis distributions agree well with the targeted topological polarization vortex configurations. Under linear polarization analysis, eight nodes appear in the polarization-analyzed far-field patterns of Figs. \ref{Fig:3}\textbf{d} and \textbf{e}, rotating in the opposite and same directions to the linear polarizer for the cases of topological charges $q=-4$ and $q=+4$, respectively, in agreement with the expected polarization distributions.

\begin{figure}[H]
 \centering
  \includegraphics[scale=1]{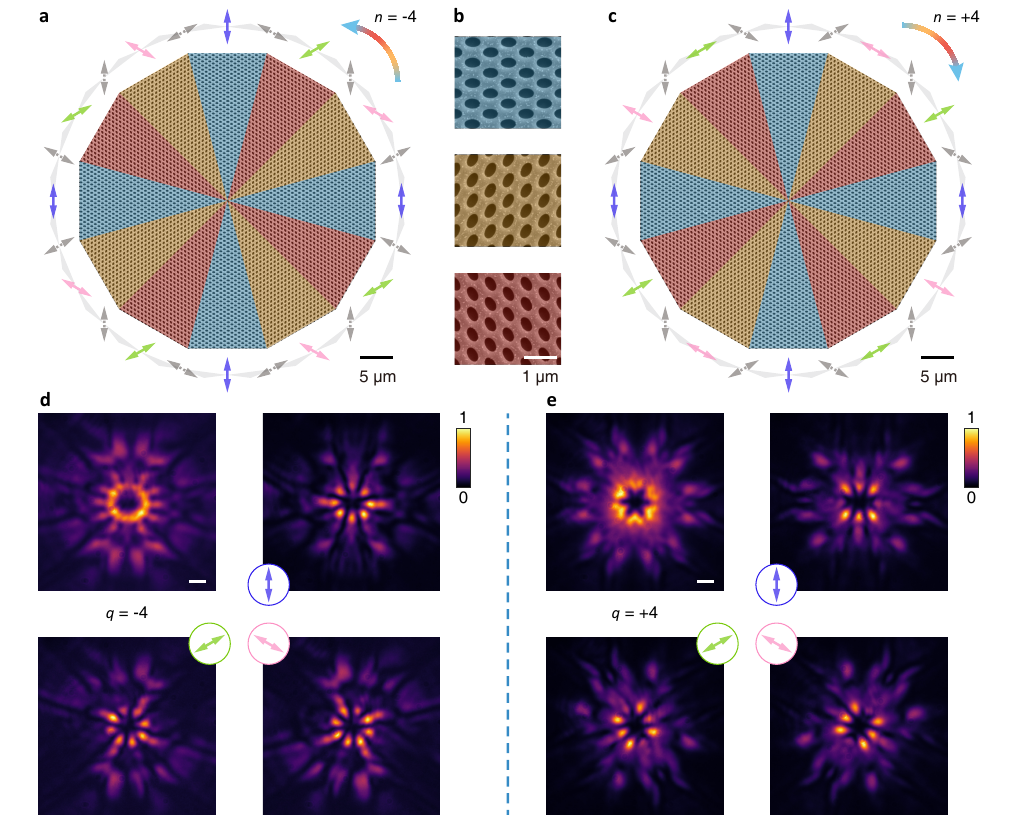}
  \caption{$\textbar$ \textbf{Experimental realization of the defect cavities for corresponding vectorial lasing.} \textbf{a-c}, SEM images of the fabricated compound cavities for vectorial lasing with topological charge of $-4$ (\textbf{a}) and $+4$ (\textbf{c}). The colored arrows outside indicate polarization states of the composed q-BIC PhC slabs. Detailed SEM images of the composed q-BIC PhC slabs are shown in \textbf{b}, with different colors highlighting the orientation of the elliptical holes. \textbf{d-e}, Measured lasing images via the corresponding compound cavities. The colored arrows show the analyzed directions of applied linear polarizers. Scale bar, $1^{\circ}$.}
  \label{Fig:4}
\end{figure}

Following the structural design in Figs. \ref{Fig:3}\textbf{b} and \textbf{c}, we experimentally fabricated the compound cavities to realize corresponding vectorial lasing. Figs. \ref{Fig:4}\textbf{a} and \textbf{c} show the SEM images of the fabricated structures supporting vectorial lasing with topological charge of $-4$ and $+4$, respectively, where different colors highlight the rotation orientations of the elliptic holes in q-BIC PhC slabs. The color definitions on the structures are detailed in Fig. \ref{Fig:4}\textbf{b}, consistent with the schematic representation in Fig. \ref{Fig:3}\textbf{a}. With optical pumping focused on the structure’s center, we experimentally realized the vectorial lasing with targeted topological charge. Figs. \ref{Fig:4}\textbf{d} and \textbf{e} exhibit the measured lasing profiles in far field. The polarization-analyzed images directly reveal the topological charges of lasing profiles, showing agreement with the simulated far-field profiles in Figs. \ref{Fig:3}\textbf{d} and \textbf{e}.

\begin{figure}[H]
 \centering
  \includegraphics[scale=1]{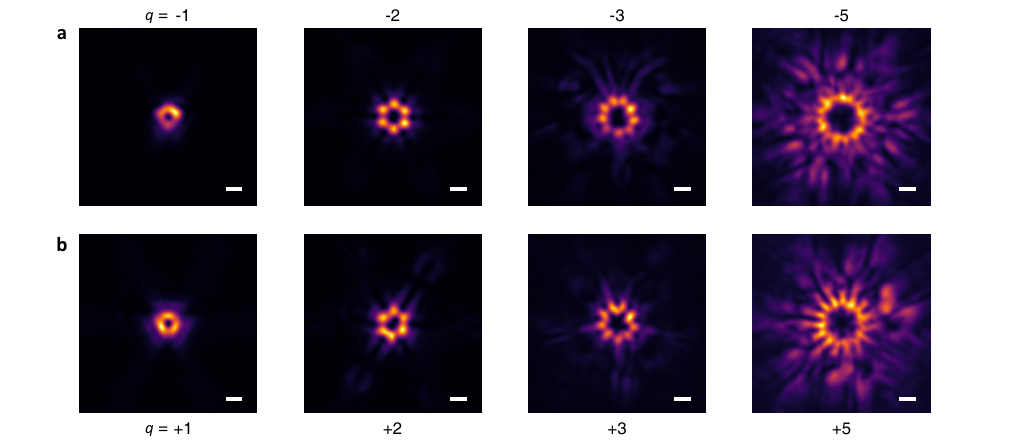}
  \caption{$\textbar$ \textbf{Experimental realization of various vectorial lasing with different topological charges.} \textbf{a-b}, Measured lasing images with  topological charges ranging from $-1$ to $-3$, $-5$ (\textbf{a}) and $+1$ to $+3$, $+5$ (\textbf{b}). The bright halos at the center correspond to vectorial lasing beams with different topological charges. Scale bar, $1^{\circ}$.}
  \label{Fig:5}
\end{figure}

Finally, Fig. \ref{Fig:5} further expands on our findings, showcasing the experimental realization of various vectorial microlasers with different topological charges based on the proposed method. Figs. \ref{Fig:5}\textbf{a} and \textbf{b} present measured total lasing profiles with topological charges ranging from $-1$ to $-3$, $-5$, $+1$ to $+3$, and $+5$, reinforcing the controllability and designability of our approach to generate vectorial lasing with a range of topological charges. Detailed structural design is presented in Supplementary Information Section 6.

\section*{Discussion}

Our work introduces a new paradigm for realizing vectorial microlasers with designable topological charges by exploiting the M\"{o}bius-like correspondence in q-BICs. This approach is fundamentally distinct from conventional BIC-based lasing, where vectorial lasing relies on the intrinsic topological vortex configurations of symmetry-protected BICs—a mechanism inherently constrained by structural symmetries~\cite{hsu2016bound,wang2024optical}, thus limiting the designability of topological charges (see Supplementary Information Section 8). In contrast, we propose first breaking the original structural symmetry to obtain PhC slabs supporting q-BICs, and then constructing new compound cavities using these q-BIC PhC slabs. The absolute value and sign of the lasing topological charges are determined by the repetition times and the direction of the angular arrangement of the q-BIC PhC sectors. Distinct from previous topological cavity designs for vectorial lasing, such as Dirac-vortex cavity, this methodology establishes a one-to-one correspondence between cavity design and lasing topological charge, offering unprecedented predictability in the design of vectorial lasing modes. More discussions on differences between Dirac-vortex cavity and the compound cavity in this work are provided in Supplementary Information Section 9.

In conclusion, we have introduced new topological concept of M\"{o}bius-like correspondence in q-BICs to construct the compound laser cavities supporting vectorial lasing with designable topological charges. Under symmetry breaking, the q-BIC with linear polarization evolves from the original high-order BIC with $-2$ topological charge in the triangular PhC slab, maintaining the high-Q property that favors lasing realization. Leveraging the unique properties of the M\"{o}bius-like correspondence, we propose a method for controlling topological charges of the compound cavity to achieve vectorial microlaser. The designability of our method was confirmed through experimentally realized vectorial lasing with topological charges ranging from $-5$ to $+5$. Our work enables purposefully engineering lasing topological charges on an ultra-compact platform. Inspired by this work, future developments can explore generating other structured light profiles and applications of vectorial lasing in advanced optical communication, sensing, imaging, and next-generation compact photonic devices that leverage topological properties for enhanced performance.

\newpage

\section*{Methods}
\noindent \textbf{Simulations.} Finite-element method (FEM) simulations were performed using COMSOL Multiphysics to obtain the photonic band structure, corresponding polarization distributions and modal profile of the lasing mode. For the simulations of the photonic band structure and corresponding polarization distributions, periodic boundary conditions were applied in the x and y directions, while the second-order scattering boundary condition was applied in the z direction. For the modal profile simulation, finite-size structure with the second-order scattering boundary condition is applied to obtain the near field ($\lvert E \rvert^2$) of the lasing mode. The calculated in-plane electric fields ($E_x$ and $E_y$ fields) of the lasing mode are Fourier-transformed to obtain the far-field profiles.

\noindent \textbf{Sample fabrication.} 
The samples were fabricated on an optical silica substrate with a 140-nm-thick Si$_3$N$_4$ layer, deposited via a plasma-enhanced chemical vapor deposition (PECVD) system (Oxford PlasmaPro System 100). Standard nanofabrication techniques were utilized. A 320-nm-thick layer of CSAR 62 positive electron beam resist was spin-coated onto the Si$_3$N$_4$ layer, followed by a conductive polymer layer (AR-PC 5092.02). Electron beam lithography (JEOL JBX-8100FS) was performed on the PhC slabs to pattern the sample. The unexposed resist served as a mask for subsequent reactive ion etching (RIE, Trion T2) using CHF$_3$ and O$_2$. After etching, the remaining resist mask was removed via RIE with O$_2$ plasma. Please see the Supplementary Information Section 1 for the schematic of the fabrication processes in this work.

\noindent \textbf{Optical measurements.} 
We implemented a momentum-space spectroscopy measurement system to carry out the optical measurements in the manuscript. Based on Fourier optics principles, the system enables analysis of momentum-space information and operates in two modes: spectrometer mode and imaging mode. In spectrometer mode, a spectrometer placed at the Fourier plane of the sample records in-plane momentum- and wavelength-resolved transmittance spectra. In imaging mode, a camera positioned at the Fourier plane captures laser beam profiles.

For the transmittance spectra measurements in spectrometer mode, we used a broadband white light source as the incident illumination. PL characterization was performed using a pulsed laser (wavelength: 800 nm; pulse width: $\sim$100 fs; repetition rate: 1 kHz) to excite samples embedded with the gain medium IR-140-DMSO (5.3 mM concentration) at ambient temperature. These measurements were also conducted in spectrometer mode. By switching to imaging mode, we recorded the beam profiles of the microlaser using a camera, with a linear polarizer introduced for polarization-analyzed measurements. For a schematic view and detailed descriptions of the optical setup, please refer to Supplementary Information Section 2.

\section*{Data availability}
We declare that all data needed to evaluate the conclusions are present in the paper and the Supplementary Information.

\section*{Acknowledgements}
This work was supported by National Key R\&D Program of China (No. 2023YFA1406900 and No. 2022YFA1404800); National Natural Science Foundation of China (No. 12234007, No. 12321161645, No. 12221004, and No. 12404427); Major Program of National Natural Science Foundation of China (Grants No. T2394480, No. T2394481); Science and Technology Commission of Shanghai Municipality (22142200400, 21DZ1101500, 2019SHZDZX01, 23DZ2260100, and 24YF2702400). J.W. was further supported by China National Postdoctoral Program for Innovative Talents (BX20230079) and China Postdoctoral Science Foundation (2023M740721).

\section*{Author contributions}
J.W., L.S. and J.Z. conceived the basic idea and designed the experiments. J.W., X.W. and Z.W. gave the theoretical explanation and performed numerical simulations. X.W. fabricated samples. X.W. and J.W. constructed the measurement systems. X.W., Z.W and J.W. performed the optical measurements. All authors analyzed the data. J.W. and X. W. wrote the manuscript, and all authors took part in the discussions and revisions and approved the final copy of the manuscript.

\section*{Competing interests}
The authors declare no competing interests.

\bibliography{Ref}
\bibliographystyle{naturemag}

\newpage

\end{document}